\documentclass[10pt,journal,compsoc]{IEEEtran}
\usepackage[table]{xcolor}
\usepackage{booktabs}
\usepackage{hyperref}
\usepackage{ulem}
\usepackage{graphicx}
\usepackage{rotating}
\usepackage{setspace}
\usepackage{longtable}
\usepackage[binary-units=true]{siunitx}
\usepackage{multirow}
\usepackage{subcaption}
\usepackage[ruled,norelsize]{algorithm2e}
\usepackage{algpseudocode}
\usepackage{amsmath}
\usepackage{tabularx}
\usepackage{xspace}
\usepackage{footmisc}

\makeatletter
\newcommand{\removelatexerror}{\let\@latex@error\@gobble}
\makeatother

\begin{document}
\title{Sea: A lightweight data-placement library for Big Data scientific computing}

\author{Val\'erie Hayot-Sasson, Mathieu Dugr\'e and Tristan Glatard}

\IEEEtitleabstractindextext{%
\begin{abstract}
    The recent influx of open scientific data has contributed to the
    transitioning of scientific computing from compute intensive to data intensive.
    Whereas many Big Data
frameworks exist that minimize the cost of data transfers, few scientific
applications integrate these frameworks or adopt data-placement strategies to mitigate the costs.
Scientific applications commonly rely on well-established command-line tools that would require complete reinstrumentation
in order to incorporate existing frameworks.
We developed Sea as a means to enable data-placement strategies for scientific
applications executing on HPC clusters without the need to reinstrument
workflows. Sea leverages GNU C library interception to intercept POSIX-compliant
file system calls made by the applications. We designed a performance model and
evaluated the performance of Sea on a synthetic data-intensive application
processing a representative neuroimaging dataset (the Big Brain). Our results
demonstrate that Sea significantly improves performance, up to a factor of
3$\times$. 

\end{abstract}
}

\maketitle

\IEEEdisplaynontitleabstractindextext \IEEEpeerreviewmaketitle

\IEEEraisesectionheading{\section{Introduction}\label{sec:introduction}}
\IEEEPARstart{E}{fficient} data-placement strategies have become essential to
minimizing Big Data processing overheads. These strategies, such as data
locality and in-memory computing, improve application runtime by ensuring data
accesses occur on the most efficient available storage. Existing Big Data
frameworks, such as MapReduce~\cite{dean2008mapreduce}, Apache
Spark~\cite{zaharia2016apache} and Dask~\cite{rocklin2015dask}, improve runtime
performance through the incorporation of data-placement strategies.

Until the recent surge in publicly available scientific data, scientific
workflows have been regarded as compute intensive. As a result, standardized
scientific tools typically lack data-placement mechanisms. In contrast, efforts
have been placed on efficient storage representation, such as HDF5~\cite{hdf5}
and Zarr~\cite{zarr}, and the sharing of data, through initiatives like
DataLad~\cite{wagner2022fairly}.

Data-placement strategies can complement existing solutions by facilitating
the transfer and processing of large datasets across different
infrastructures, and placing the different file formats at the preferred
storage layer to maximize efficiency. However, adapting existing scientific
applications to Big Data frameworks requires considerable effort,
necessitating the rewrite of many well-established applications.
Furthermore, as noted in \cite{mehta2017comparative}, Big Data frameworks
are not easy to use for the processing of certain scientific data, such as
imaging data, due to the level of expertise required to rewrite reference
implementations using the frameworks.




Our goal is to provide a data-placement solution to standardized scientific
computing applications executing on High Performance Computing (HPC) clusters.
The applications of interest to us consist of various tasks that communicate
to each other via a POSIX-compliant file system. Whereas the application can be
platform-agnostic, we assume users will leverage them within an linux-based HPC context,
particularly when processing large volumes of data, due to their inherent
reliance on POSIX-compliance.

We developed Sea, a user-space data-placement library that provides efficient
data placement without reinstrumentation of the underlying application. Given a
predefined storage hierarchy, Sea can prefetch, cache, flush, and evict data to
and from a permanent storage location.

To provide a
frame-of-reference for Sea's performance on HPC systems, we designed a
performance model. We then evaluated Sea's performance on a synthetic Big Data
application running on a representative neuroimaging dataset, comparing its
performance with the bounds delineated by the model.

\section{Related Work}
\subsection{HPC Infrastructure}
      The general structure of HPC clusters complicates the deployment of Big
      Data frameworks. Typical HPC clusters consist of distinct storage and
      compute nodes. While compute nodes may also have local storage, there is
      no distributed file system like the Hadoop Distributed File System
      (HDFS)~\cite{shvachko2010hadoop} or Alluxio~\cite{alluxio} to facilitate
      data locality. Furthermore, access to these compute nodes is ephemeral,
      with allocation duration enforced by a batch scheduler. Therefore, any
      data written to compute nodes are rendered inaccessible after the
      allocation is terminated.

      The storage layer found on HPC clusters consists of a high-performance
      parallel file system (PFS) (e.g., Lustre~\cite{lustre}). These file systems
      are shared amongst all compute nodes within a cluster, typically connected
      to the nodes via high-performance network interconnect like
      InfiniBand~\cite{infiniband}. I/O overheads of shared PFS can be incurred
      in many places, from the shared network to the bandwidth and latency of
      the storage devices.  With Lustre specifically, there typically are many
      data nodes, known as Object Storage Servers (OSS) which contain several
      storage devices, known as Object Storage Targets (OST). All file metadata
      is maintained within a separate node known as the Metadata Server (MDS)
      and stored within a device referred to as the Metadata Target (MDT).
      Although data transfers can be communicated directly to the corresponding
      OST, the clients need to first communicate with the MDS to determine which
      OST to communicate with. This can result in major data transfer overheads,
      particularly when making numerous requests to the metadata server.

      HPC clusters may provide a faster intermediate storage layer, known as a
      Burst Buffer, aimed at improving I/O performance. A Burst Buffer typically
      consists of higher-performance storage devices (SSD and memory) and can be
      local to the compute node or as a dedicated I/O node. Burst Buffers were
      introduced as a response to offset the impacts of large-scale application
      checkpointing on the PFS~\cite{bb}. The application would write a
      checkpoint to the Burst Buffer and resume processing, while the checkpoint
      would be asynchronously flushed to the PFS.

\subsection{Big Data Frameworks and File systems}

Early Big Data
      frameworks, notably MapReduce~\cite{dean2008mapreduce} and Apache
      Spark~\cite{zaharia2016apache} have implemented and popularized
      data-management strategies to minimize data transfers at processing time.
      These strategies are known as data locality and in-memory computing.
      Data locality ensures that compute tasks are scheduled nearest to where
      the data is located. Historically, storage and compute nodes were distinct
      layers, transferring data to the compute location whenever necessary. Rather than transferring
      large amounts of data over the network to compute tasks, which could incur
      significant overheads, Big Data frameworks ensure that data is stored
      directly on the compute nodes. When a compute task requires access to
      specific data, the scheduler sends the task to the nearest available node
      to the data, thereby minimizing any cost of network-related data
      transfers. This strategy is not only used by Big Data Frameworks such as
      Hadoop MapReduce, Apache Spark, and Dask~\cite{rocklin2015dask}, but also
      enabled by file systems such as HDFS and Alluxio.

      In-memory computing complements data locality by maximizing use of
      available memory to maintain intermediate data. Whereas with data locality
      alone, tasks would leverage local storage, typically in the form of HDDs
      or SSD, to maintain task input and output data, with the addition of
      in-memory computing, data would be stored in main memory to minimize
      latency and maximize bandwidth.

      While Big Data frameworks and filesystems are not commonly used in HPC,
      HPC clusters are typically Linux-based and benefit from the Linux page
      cache, which can provide data locality and in-memory computing at a
      limited capacity. Furthermore, HPC architectures may also provide an
      intermediate storage layer, known as a Burst Buffer, to improve
      application I/O-related overheads. To facilitate an application's
      interaction between application and storage layer, distributed file
      systems have been developed.

\subsection{The Linux page cache}
 While scientific applications do not customarily leverage data-placement
      strategies and may perform large over-the-network data transfers, they may
      still benefit from in-memory computing and data locality through the Linux
      Page Cache~\cite{pagecache}. Similarly to other file systems, Lustre
      leverages the compute node page cache to reduce I/O overheads. System
      memory is composed of two components: 1) anonymous memory and 2) page
      cache. Anonymous memory consists of all application-related objects,
      whereas page cache consists of recently accessed file data. When a file is
      read, that file is loaded up into the page cache to be flagged for
      eviction based on a least recently used (LRU) policy. That means
      subsequent accesses to that data may be done entirely in memory so long as
      the data has not already been evicted. Similarly, for writes to a file
      system with writeback cache enabled, the file will be written to memory
      completing the write operation once the file has been written entirely to
      memory. That file will then be flushed asynchronously to the appropriate
      storage device. As system memory may get overloaded with too many write
      requests, there is a limit to the amount of written data that can exist in
      memory, known as the dirty\_ratio. Furthermore, applications producing too
      many write requests may be throttled by the system.

\subsection{File System implementations}
      Due to the architecture of conventional HPC clusters, network-based
      parallel files systems are favoured over Big Data distributed file
      systems. Such file systems require super-user access, preventing users
      from deploying their own on-the-fly cluster. While HDFS can be loaded in
      user-space by mounting its FUSE (Filesystem in User Space) implementation,
      FUSE-based file systems may perform significantly worse than desired,
      depending on the application~\cite{tofuse}. Even with a user-space version
      of HDFS, there would be no mechanism to transfer all the data back to the
      PFS post-processing.
      
      However, there exist Big Data file systems, such as
      XtreemFS~\cite{xtreemfs}, that exist entirely within user space, are
      POSIX-compliant, and can use alternative methods to FUSE to function. For
      instance, XtreemFS uses the LD\_PRELOAD trick to intercept file system
      calls made to the GNU C library (glibc). There are limitations to using
      glibc interception; the application must be dynamically-linked and make
      glibc calls. However, statically-linked applications are uncommon and
      applications interacting with POSIX-compliant file systems on Linux
      machines, the predominant OS of HPC clusters, will make glibc calls.
      Nevertheless, alternatives to glibc intercept exist that can bypass
      the aforementioned issues, such as system call interception, but they
      result in greater overheads~\cite{quinson}.

      Due to complex file system deployment and configuration properties,
      XtreemFS is unlikely to be deployed on an ephemeral burst buffer. It also
      does not provide a simple way to leverage different classes of storage
      devices, nor ensure that required data will be copied to the cluster's
      parallel file system.

      BurstFS~\cite{burstfs} and GekkoFS~\cite{gekkofs} are two user-space file
      systems that are specifically designed with HPC Burst Buffers. That is, both of
      these file systems are designed to support the ephemeral nature of HPC compute nodes and their associated storage.
      Additionally, these implementations exist entirely in user-space leveraging the LD\_PRELOAD trick to intercept
      necessary library calls. Whereas these libraries incorporate a client-server architecture to aid in transferring data
      between nodes, Sea opts for a more lightweight approach ensuring decentralization, statelessness and leveraging underlying
      file systems for communication and consistency.

\section{Materials and Methods}

\subsection{Sea design and implementation}

\begin{figure*}

    \centering
    \includegraphics[width=\columnwidth]{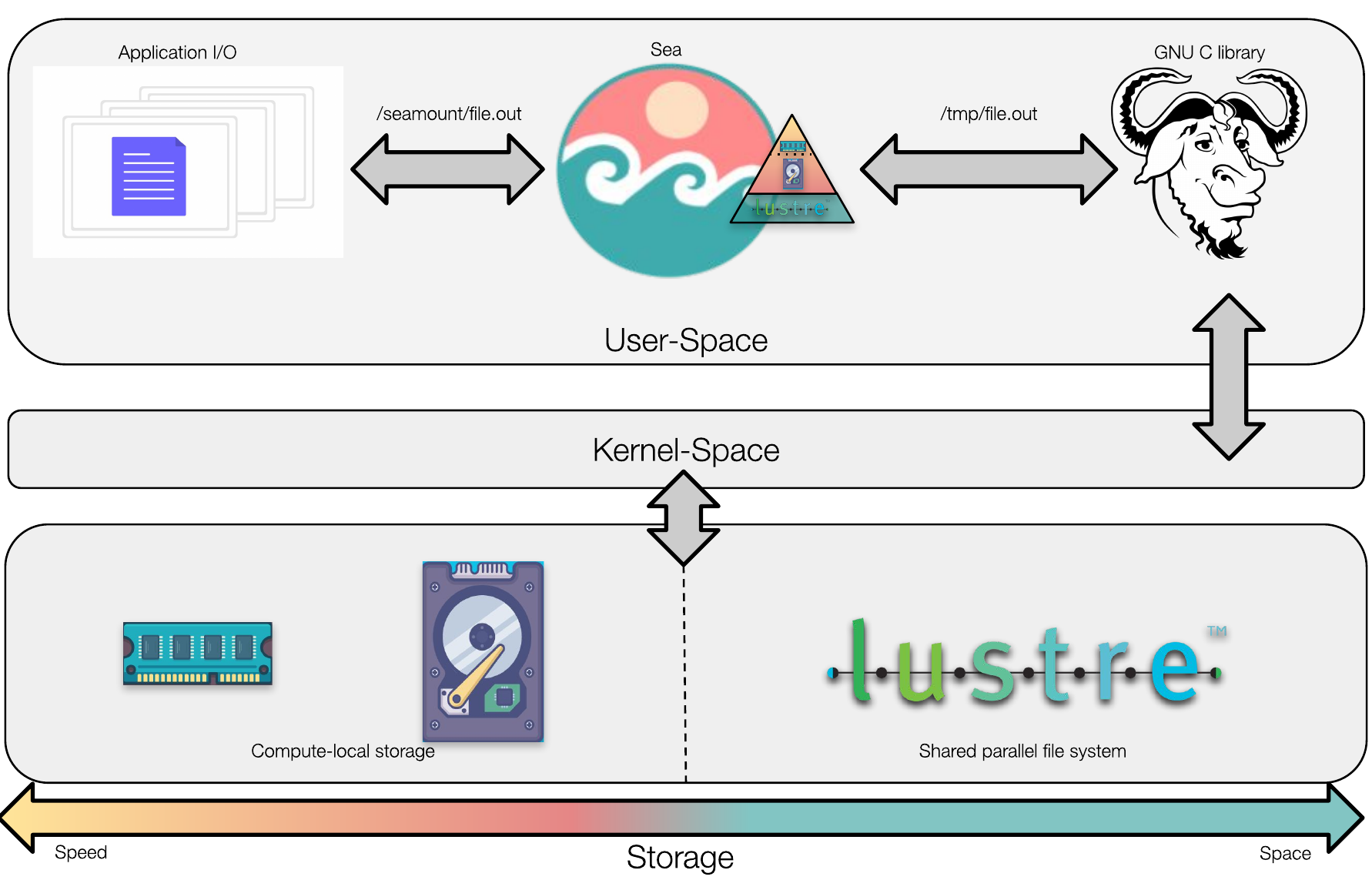}%
\caption{The Sea data-placement library. Sea intercepts I/O-related glibc calls
issued by the application and translates the path provided if located within the
Sea mountpoint. Sea leverages a user-defined storage hierarchy to help select
the most appropriate path and the redirects the call to glibc to process the
request on the translated path. }
\label{fig:sea-comp:diagram}
\end{figure*}

Sea\footnote{\url{https://github.com/big-data-lab-team/sea}} is an open-source
user-space data-placement C++ library for applications running on HPC systems.
Its main aim is to reduce application data transfer costs by leveraging
node-local storage (Figure~\ref{fig:sea-comp:diagram}). Sea redirects files
accessed from a user-specified mount point to the appropriate storage devices
using glibc interception. Within an intercepted call the file will be written to
or read from the fastest storage device available.

Sea requires minimal configuration for use. In order to work, Sea requires the
user to specify at least two storage devices, a fast temporary device used as cache
and a slower long-term storage device. This could be RAM and SSD if working on a
single node, or a compute-local SSD and a shared PFS, in the
case of an HPC cluster. Ideally, a user will provide a multitude of ephemeral
storage devices to improve Sea's efficacy. Furthermore, to maximize usage of the
fastest available storage devices, Sea will allow the user to outline which
files can be removed from short-term storage in addition to which files need to
be materialized onto long-term storage. Eviction of files is important as this
will allow Sea to maximize usage of short-term storage.

Sea was created as a data-placement service to be employed at the discretion of
the users on scientific applications. When developing the idea for Sea, we
observed the design differences between Big Data and scientific, specifically
neuroimaging, frameworks and applications. Scientific frameworks and
applications tend to favour ease-of-use, reproducibility, portability and
parallelism over minimizing data
transfers\cite{jarecka2020pydra,esteban2019fmriprep,wagner2022fairly}. It was
important for us to ensure that Sea would preserve the design decisions of the
underlying tools. Sea achieves this by having few dependencies, is lightweight,
and requires minimal configuration.

In this subsection, we will discuss and outline the various design and
implementation details made. Furthermore, we will go into detail into the
functionality of Sea.

\subsubsection{Requirements and assumptions}

Sea removes the need to implement the logic to redirect data to the appropriate
storage locations. Furthermore, it implements logic to ensure that files are
written to the best possible location at any given time. It is both pipeline and
infrastructure agnostic.

As it executes in user-space, root permissions are not required, enabling users
to use Sea on most Linux-based systems available to them. This differs from Big
Data file systems which may require elevated permissions to install. The
overhead of intercepting glibc calls is minimal, and negligible compared to
system call interception and file systems such as FUSE.

Unlike Big Data frameworks, Sea does not require reinstrumentation of the
existing pipelines, allowing users to gain an instant performance boost.

Sea is primarily designed to work with workloads that generate large amounts of
intermediate data. Since data must be read from the slower storage device and final outputs will
also have to be written to the same device, the
use of Sea with workloads that do not produce intermediate data would result in
limited speedup and may even introduce some overheads. 

Sea provides two main modes based on flushing specification: In-memory computing
and copy-all. In-memory computing is achieved when intermediate need not be
materialized to long-term storage (i.e. no flushing). With Sea in-memory, we can
expect speedups comparable to Big Data frameworks.

As the name implies, Sea copy-all is applicable when all data must be
materialized to long-term storage (i.e. flush everything). This can be coupled
with eviction to limit the amount of data that will be written to slower, but
larger, local storage devices. While this option will have overheads, due to the
need to flush all application data, overheads will be negligible when data transfer and compute time
are similar. 


One important assumption that must be made is that the amount of data produced
by the workload far exceeds the amount of page cache space available and
utilized by the different file systems. In the case where all workload data can
fit into page cache memory, the benefit of Sea may be greatly diminished if not
negligible.

Naturally, the amount of data written may not fit entirely into a single local
storage location. Currently, the user cannot specify the maximum amount of space
to use in a given mountpoint explicitly. Sea queries all the available file
systems directly to determine the amount of available space. It relies on
user-specified information on the maximum file size produced by the pipeline and
the number of concurrent pipeline parallel threads to determine if there is
sufficient space to write to the file system without issue. While it prevents
overheads due to locking, it makes the assumption that the user is alone writing
to the storage device. The user can reduce the amount of local storage space
used by increasing the maximum file size of the application.


\subsubsection{Glibc interception}

Glibc interception is achieved by writing wrappers to existing glibc functions.
Most importantly, every glibc function accepting a file path as input parameter needs to be
wrapped. The wrappers take any input filepath that is located within the
user-provided Sea mountpoint and convert it to a filepath pointing to the best
available storage device.

Sea does not currently support the partitioning of files across multiple
devices. Since it cannot predict the size of the outputs to ensure the existence
of sufficient space on storage devices, the user must provide within the Sea
configuration file the maximum file size produced by the workflow. Together with
the specified amount of parallel processes, Sea calculates the minimum space
required on a storage device to write the file to it.
Sea will then go through the hierarchy of available storage devices and select
the fastest storage device with sufficient available space. While this process
may result in not utilizing all available space in fast storage devices, it will
still allow users to gain a performance speedup on the pipeline given that the
number of threads multiplied by the file size does not exceed storage space.

\subsection{Testing}
To work effectively, Sea has to intercept every glibc function that interacts
with files or directories. Failure to intercept some of these function may
result in the whole application crashing. This is because only Sea can translate
the Sea mountpoint file paths to their respective real locations. Due to an
absence of standardized glibc testing suites, we developed our own integration
tests.

Our tests have been executed on three different operating systems: Ubuntu
(xenial, bionic, focal), Fedora (32, 33, 34) and Centos (6, 7). Using this test
matrix, the Sea tests were able to span seven glibc versions ranging from 2.17
to 2.33. As glibc file system functions may change over time, Sea behaviour with
newer and older versions of glibc is unknown. 

\subsection{Memory-management}
      \begin{table}
      \centering
      \begin{tabular}{ccc}
        \toprule
       Mode & .sea\_flushlist & .sea\_evictlist \\
       \midrule
       Copy & yes & no \\
       Remove & no & yes \\
       Move & yes & yes \\
       Keep & no & no \\

       \bottomrule

      \end{tabular}
      \caption{Sea's memory management modes}
      \label{table:sea-comp:fe}
      \end{table}
Memory-management is an integral component of Sea as de-allocating data from cache
makes room for future files to be stored within the higher-performance storage devices.
Generalizing an efficient memory-management strategy across all applications is a challenging task.
Since Sea works on a per-application basis, memory-management in Sea is application-specific, configured
via two files: \texttt{.sea\_flushlist} and \texttt{.sea\_evictlist}.

Memory management in Sea consists of four different modes, as described in Table~\ref{table:sea-comp:fe}: copy,
remove, move and keep. A copy copies a file located within the Sea cache to
persistent storage. This operation is particularly useful when a file will be
reused by the application, and thus benefits from remaining in cache, but also
needs to be communicated between other nodes or needs to be persisted for
post-processing analysis. To copy a file, the file must only be specified
in the \texttt{.sea\_flushlist}.

Remove is for files that are
located within a Sea cache, but do not need to be persisted and will not be
reused by the pipeline and therefore do not need to occupy cache space.
Remove is useful for removing unnecessary application log files. To
remove a file, the file must only be specified in the
\texttt{.sea\_evictlist}.

Similar to the copy, the move mode copies data from cache to
persistent storage. However, the assumption here is that these files will not be
reused by the application and therefore need not occupy space in cache.
Move is therefore synonymous with a copy-and-remove operation and is invoked on
files that are specified in both the \texttt{.sea\_flushlist} and
\texttt{.sea\_evictlist}.

Keep is useful for when file needs to be in cache as it will be reused by the
application, but is not needed in post-processing analysis or by other nodes.

A third text file, namely the \texttt{.sea\_prefetchlist}, exists that can be populated by input
files to be prefetched. At this time, Sea cannot determine when prefetched files are no longer needed by the pipeline, and
therefore should not be evicted. For files to be prefetched, they must be located within Sea's mountpoint at startup.


\subsection{The Sea and Lustre model}\label{ss:sea-comp:model}

      Experimental results can be prone to error and variability, particularly
      in instances where there a many factors at play, such as in distributed cluster
      environments. To improve confidence in our experimental results, we developed a
      simplified performance model that could help describe the behaviours observed in our experimental
      results. Since different PFS may
      operate differently, our baseline model will be based on Lustre which is
      commonly used on HPC infrastructure.

      For data-intensive use cases, the makespan models for both Lustre and Sea
      can be broken down into two components: The amount of time it takes read
      the data and the amount of time it takes to write the data. With more
      heterogeneous applications (some components are compute-intensive whereas
      others are data-intensive), a third component, consisting of compute time,
      can be added. Furthermore, latency may also play a significant role in an
      application makespan, particularly in scenarios with large amounts of
      small files. We choose to ignore latency costs in our model and make the
      assumption that the application bottleneck is the bandwidth, however, for
      more accurate estimates we might consider the addition of file system
      latency, as a fourth model component.

      A simplified version of the Lustre makespan model can be defined as
      follows:

      \begin{equation}\label{eq:sea-comp:lustrenpc}
          M_{l} =  \frac{D_{r}}{L_{r}} + \frac{D_{w}}{L_{w}}
      \end{equation}

      {\noindent} Where, \\
      $M_{l}$ is the application makespan using Lustre \\
      $D_{r}$ is the amount of data read \\
      $D_{w}$ is the amount of data written \\
      $L_{r}$ is Lustre's read bandwidth \\
      $L_{w}$ is Lustre's write bandwidth \\



      To determine the Lustre bandwidth, one must consider the three components
  involved: 1) the network bandwidth of the compute nodes, 2) the network
  bandwidth of the data nodes, and 3) the collective bandwidth of the Lustre
  storage devices. Depending on each component's respective values, either of
  the three may be the source of a bottleneck. The Lustre bandwidth read and
  write models can therefore be described as follows:

    \begin{equation} 
        L_{r} = \min{(cN, sN, d_{r}\min{(d, cp)})}
    \end{equation}

    and


    \begin{equation}
        L_{w} = \min{(cN, sN, d_{w}\min{(d, cp)})}
    \end{equation}

    {\noindent} Where, \\
    $N$ is the network bandwidth \\
    $c$ is the number of compute nodes used \\
    $s$ is the number of Lustre storage nodes \\
    $p$ is the number of parallel application processes \\
    $d$ is the number of Lustre storage disks \\
    $d_{r}$ is the read bandwidth of a single Lustre storage disk \\
    $d_{w}$ is the write bandwidth of a single Lustre storage disk \\




      For the sake of simplicity, the above models assume that the network
      bandwidth between the compute and data nodes is the same. However, this
      may not necessarily be the case. Furthermore, the model also assumes that
      each file can only be located on a single disk, meaning that the parallel
      bandwidth can at maximum be as fast as all Lustre disks combined and as
      slow as the minimum number of compute threads reading and writing files.

      As with many file systems, page cache plays an important role in the speed
      of application read and writes in Lustre. Since the effects of page cache
      may be non-negligible given amount of memory available and the data
      accessed during the execution of the application, it is important to
      include it in our model. The makespan of an application I/O to and from
      page cache, $M_{c}$, can be described as in
      Equation~\ref{eq:sea-comp:lustrenpc}, where it is assumed that none of the
      data is written or read from page cache.

      \begin{equation}\label{eq:sea-comp:cache}
          M_{c} = \frac{D_{cr}}{cC_{r}} + \frac{D_{cw}}{cC_{w}}
      \end{equation}

      {\noindent} Where, \\
      $M_{c}$ is the makespan of application I/O to page cache \\
      $D_{cr}$ is the amount of data read from cache \\
      $D_{cw}$ is the amount of data written to cache \\
      $C_{r}$ is the page cache read bandwidth \\
      $C_{w}$ is the page cache write bandwidth \\ 



       As each individual compute node has its own set of memory, we treat the
      total memory bandwidth as the sum of the individual memory bandwidth of
      each compute node.

      Page cache is difficult to summarize accurately within a
      model. For one, we must not only consider available memory and anonymous
      memory used by the application, but we must also consider which pages are
      candidates for eviction and which files they belong to. In addition, in
      the case of writes, we must consider asynchronous flushing and the
      throttling that may occur as a consequence of surpassing the
      dirty\_ratio. Furthermore, Lustre also has its own user-defined
      settings for how it interacts with the cache that would add additional
      complexities to the model. As a result, we assume two possible scenarios,
      one in which page cache is never used
      (Equation~\ref{eq:sea-comp:lustrenpc}) and one in which all application
      I/O occurs entirely within page cache, excluding the first read which must
      occur on Lustre (Equation~\ref{eq:sea-comp:lustrepc}). These two models
      allow us to define the bounds of Lustre's performance.

      \begin{equation}\label{eq:sea-comp:lustrepc}
          M_{lc} = \frac{D_{I}}{L_{r}} + M_{c}
      \end{equation}

      {\noindent} Where, \\
      $M_{lc}$ is the application makespan using Lustre with page cache \\
      $D_{I}$ is the amount of input data \\
      $L_{r}$ is the Lustre read bandwidth \\
      $M_{c}$ is the application makespan on page cache \\



      Sea's model is more complex than Lustre's as there can be several layers
      of different devices. For instance, Sea's model can be defined as:

      \begin{equation}\label{eq:sea-comp:sea}
          M_{S} = \frac{D_{I}}{L_{r}} + M_{1} + \cdots + M_{n}
      \end{equation}

      Here, $M_{n}$ represents the makespans of the different possible storage
      levels (e.g., tmpfs, NVMe, SSD, HDD, Lustre) and $M_{S}$ represent the
      makespan of an application reading and writing to Sea. For our model, we
      will assume three storage layers: 1) fast tmpfs, 2) intermediate local SSD
      storage, and 3) slow parallel file system layer. It is important to note that
      the model assumes that I/O never occurs in distinct storage layers in parallel.
      Although Sea tries to exhaust the layers, starting from the fastest, it is nevertheless
      possible that Sea writes to two storage layers at the same time, for instance, when one becomes full
      and another parallel thread needs to perform a write.

      Since the modelling of page cache is even more challenging with Sea due to
      the additional tmpfs and SSD layer, we will will model the upper and lower
      performance bounds, as we did with Lustre. Using the three layers and
      disregarding any possible effects of caching, we can redefine the Sea
      model to be:

      \begin{equation}\label{eq:sea-comp:snc}
          M_{S} = M_{SL} + M_{Sg} + M_{St}
      \end{equation}

      Where, \\
      $M_{S}$ is the makespan of an application using Sea \\
      $M_{SL}$ is the makespan of the Sea I/O to Lustre \\
      $M_{Sg}$ is the makespan of the Sea I/O to local disks \\
      $M_{St}$ is the makespan of the Sea I/O to tmpfs \\


      The tmpfs component of the Sea makespan can be defined as the amount of
      data that can be written ($D_{tw}$) to and read ($D_{tr}$) from tmpfs over
      its respective bandwidths ($C_{r}$ and $C_{w}$). In other words:

      \begin{equation}\label{eq:sea-comp:mst}
          M_{St} = \frac{D_{tr}}{cC_{r}} + \frac{D_{tw}}{cC_{w}}
      \end{equation}

      \begin{equation*}\label{eq:sea-comp:dtr}
          D_{tr} = \min\left(D_{m}, \max{\left(c(t - pF), 0 \right)} \right)
      \end{equation*}
      \begin{equation*}\label{eq:sea-comp:dtw}
          D_{tw} = \min\left(D_{m} + D_{f}, \max{\left(c(t - pF), 0 \right)} \right)
      \end{equation*}

      Where,\\
      $D_{tr}$ is the amount of data read from tmpfs \\
      $D_{tw}$ is the amount of data written to tmpfs \\
      $D_{m}$ is the amount of intermediate data \\
      $D_{f}$ is the amount of final output data \\
      $t$ is the amount of available space on tmpfs \\
      $F$ is the size of a single file \\

      In an optimal scenario all intermediate data and final output data would
      fit in tmpfs. This would provide an application using Sea in-memory
      performance. However, due to limited tmpfs storage space, it is unlikely
      to be the case. In addition, Sea may further restrict available storage
      space to prevent exceeding tmpfs storage by ensuring that there is at
      least sufficient space for all processes to each write a file
      concurrently.

      The local disk  makespan model is similar to the tmpfs makespan model,
      although we must disregard any data that has already been
      written to tmpfs. Furthermore, in Sea, it is possible to leverage however
      many disk-based file systems are available for use ($g$). For our model,
      we assume that the size of each device is identical. The makespan model
      can be defined as follows:

      \begin{equation}\label{eq:sea-comp:msd}
          M_{Sg} =  \frac{D_{gr}}{gcG_{r}} + \frac{D_{gw}}{gcG_{w}}
      \end{equation}

      \begin{equation*}\label{eq:sea-comp:ddr}
          D_{gr} = \min{(D_{m} - D_{tr}, \max{(c(gr - pF),0)})}
      \end{equation*}

      \begin{equation*}\label{eq:sea-comp:ddw}
          D_{gw} = \min{(D_{m} + D_{f} - D_{tw}, \max{(c(gr - pF),0)})}
      \end{equation*}

      Where, \\
      $D_{gr}$ is the amount of input data read from local disk \\
      $D_{gw}$ is the amount of data written to local disk \\
      $g$ is the number of local disks on a compute node \\
      $r$ is the amount of disk space available on a given disk \\
      $G_{r}$ is the read bandwidth of the disks \\
      $G_{w}$ is the write bandwidth of the disks \\

      The final component of the Sea model is the Lustre component
      (Eq.~\ref{eq:sea-comp:msl}). Sea's Lustre makespan model consists of the
      initial read from Lustre and includes and data that must be written to
      Lustre due to insufficient space on local storage and the makespan to read
      the intermediate data from Lustre.

      \begin{equation}\label{eq:sea-comp:msl}
          M_{SL} = \frac{D_{I}}{L_{r}} + \frac{D_{Lr}}{L_{r}} + \frac{D_{Lw}}{L_{w}}
      \end{equation}
      \begin{equation*}\label{eq:sea-comp:dlr}
          D_{lr} = D_{m} - D_{gr} - D_{tr}
      \end{equation*}
      \begin{equation*}\label{eq:sea-comp:dlw}
          D_{lw} = D_{m} + D_{f} - D_{gw} - D_{tw}
      \end{equation*}

      Where, \\
      $D_{Lr}$ is the amount of data read to Lustre by Sea \\
      $D_{Lw}$ is the amount of data written to Lustre by Sea \\

      Sea and Lustre have an identical lower bound. That is, ideally, both must
      perform the first read from Lustre, but all subsequent data accesses can
      be performed entirely within the page cache. The page cache model for Sea
      can be defined as the following:

      \begin{equation}\label{eq:sea-comp:msc}
          M_{Sc} = \frac{D_{I}}{L_{r}} + \frac{D_{m}}{cC_{r}} + \frac{D_{m} + D_{f}}{cC_{w}}
      \end{equation}
\subsection{Experiments}
\subsubsection{Application}

      To evaluate the performance of Sea with data intensive applications, we wrote a simple Python
      application based off of Algorithm~\ref{alg:sea-comp:incrementation}.
      
      Using this application, we can easily control how much intermediate data
      is produced by altering the amount of iterations required. Although our
      model should be able to support images of different sizes, we wanted to
      minimize any possible scheduling effects from our experiments. Therefore,
      each application processes the same amount of input data, produces the
      same amount of intermediate and output data, and performs the same amount
      of computation.

      \begin{figure}[!t]
        \removelatexerror
        \begin{algorithm}[H]
          \caption{Incrementation}\label{alg:sea-comp:incrementation}
          \SetAlgoLined \SetKwInOut{Input}{Input} \Input{$n$ number of
          iterations;\\
                $C$ a set of image chunks;\\
                $fs$ the file system to write to (Lustre or Sea)\\
            } \ForEach{$chunk \in C$} { read $chunk$ from Lustre \For{$i \in
            [1,n]$} { $chunk\gets chunk+1$ save $chunk$ to $fs$ } }
        \end{algorithm}
      \end{figure}

      We use the BigBrain~\cite{amunts2013bigbrain}, a one-of-a-kind histology
      dataset of the human brain, as a representative scientific dataset. For
      all our experiments, we utilize the \SI{20}{\micro\meter} dataset, which
      totals to approximately \SI{603}{\gibi\byte}. The dataset was broken down
      into 1000 files each consisting of \SI{617}{\mebi\byte} of data.

      We evaluated Sea using 4 different experimental conditions: 1) varying the
      number of nodes, 2) varying the number of disks, 3) varying the number of
      threads, and 4) varying the number of iterations. Experimental condition 1
      tests the effects of increasing concurrent accesses to Lustre while fixing
      disk parallel threads (i.e., the number of threads per node is constant).
      Condition 2 varies disk contention while fixing
      contention to Lustre, whereas condition 3 evaluates the effects of contention on both
      Lustre and local storage. Experimental condition 4 varies the total amount
      of intermediate data produced by the application. The fixed conditions for
      the experiment were 5 nodes, 6 processes, 6 disks, 10 iterations and 1000
      blocks. Sea in-memory (flushing and eviction of only the last iteration of
      files) strategy was used for all conditions.
      
      We also evaluated Sea flush-all (flush all files to Lustre) using 5 nodes,
      64 processes, 6 disks and 5 iterations using the same incrementation
      application.

\subsubsection{Infrastructure}
      \begin{table}
      \centering
      \begin{tabular}{ccc}
       \toprule
       Storage layer & Action & Avg. bandwidth (MiB/s) \\
       \midrule
      \multirow{3}{*}{tmpfs} & read & 6676.48 \\
      & cached read & 6318.08  \\
      & write & 2560.00 \\
      \midrule
       \multirow{3}{*}{local disk} & read & 501.70  \\
       & cached read & 7034.88 \\
       & write & 426.00 \\
       \midrule
       \multirow{3}{*}{Lustre} & read & 1381.14 \\
       & cached read & 6103.04  \\
       & write & 121.00  \\

       \bottomrule

      \end{tabular}
      \caption{Storage benchmarks}
      \label{table:sea-comp:fs}
      \end{table}
            Our experiments were executed on a Centos 8.1 (Linux kernel 4.18.0)
      cluster with 8 compute nodes, a 4 data node Lustre server with 1
      metadata node. Each compute node is equipped it two Intel(R) Xeon(R) Gold
      6130 CPUs, \SI{250}{\gibi\byte} of memory with \SI{126}{\gibi\byte} of
      tmpfs space and 6 \SI{447}{\gibi\byte} Intel SSDSC2KG480G8R SSDs. The data
      nodes each contain 11 \SI{10}{\tera\byte} HGST HUH721010AL HDD OSTs and
      \SI{62}{\gibi\byte} memory. The metadata server contains a Toshiba
      KPM5XVUG960G \SI{960}{\giga\byte} MDT. The network bandwidth is
      25~GbE and uses TCP for communication. Jobs are scheduled on
      the cluster from a controller node using Slurm with cgroups. Swapping is
      disabled and Lustre dirty writes is limited to \SI{1}{\giga\byte} per OST.

      There was no background load on the cluster for the duration of the experiments.

      Each file system was benchmarked 5 times with the \texttt{dd} command \footnote{\url{https://github.com/big-data-lab-team/Sea/blob/master/benchmarks/scripts/bench_disks.sh}}.
      The average bandwidths are reported in Table~\ref{table:sea-comp:fs}

\section{Results}

    \begin{figure*}

    \begin{subfigure}{\columnwidth}
        \centering
        \captionsetup{width=.85\linewidth}
        \includegraphics[width=\columnwidth]{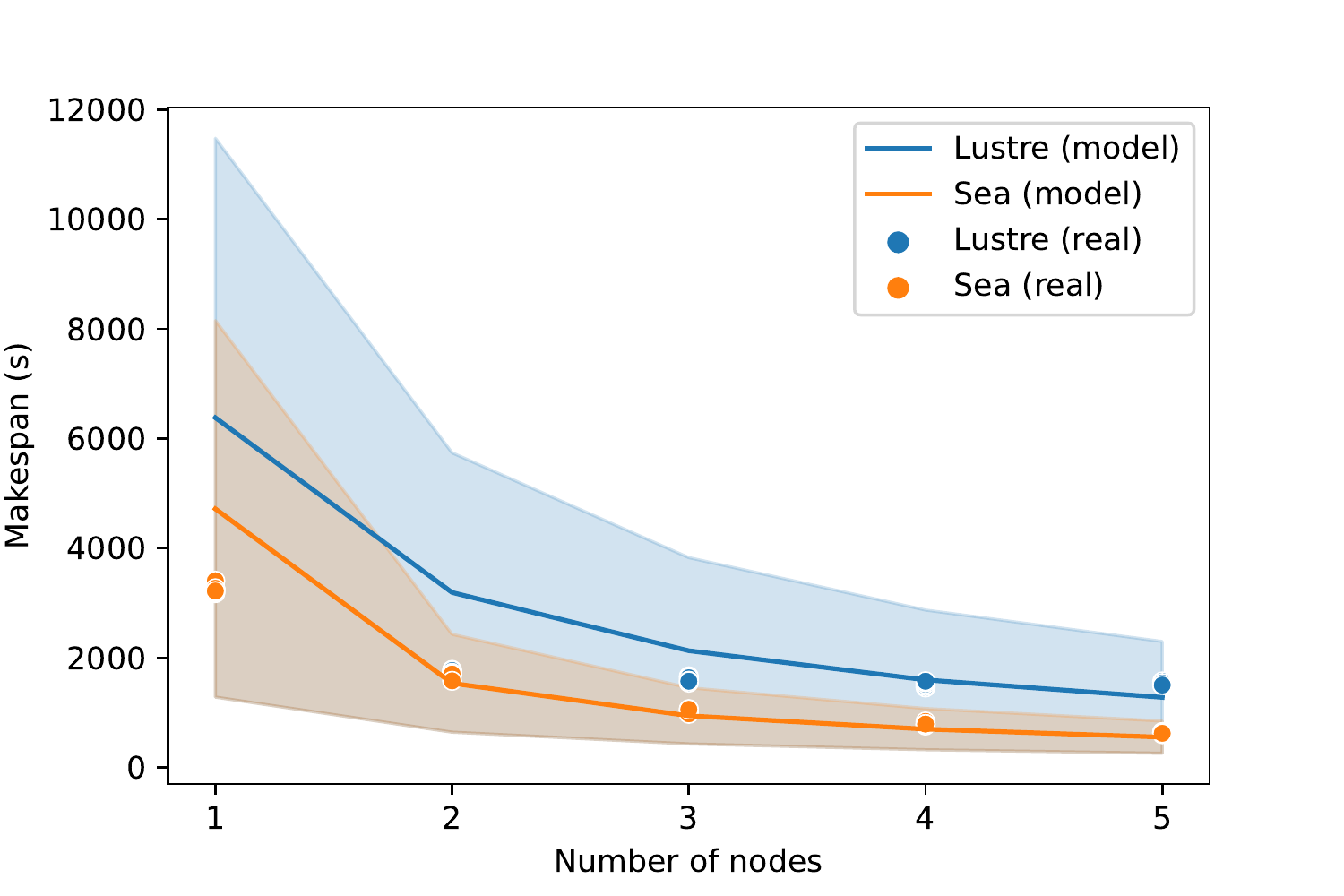}%
        \caption{Experiment 1: Varying the number of nodes, 10
        iterations}\label{fig:sea-comp:nodes}
    \end{subfigure}
    \begin{subfigure}{\columnwidth}
        \centering
        \captionsetup{width=.85\linewidth}
        \includegraphics[width=\linewidth]{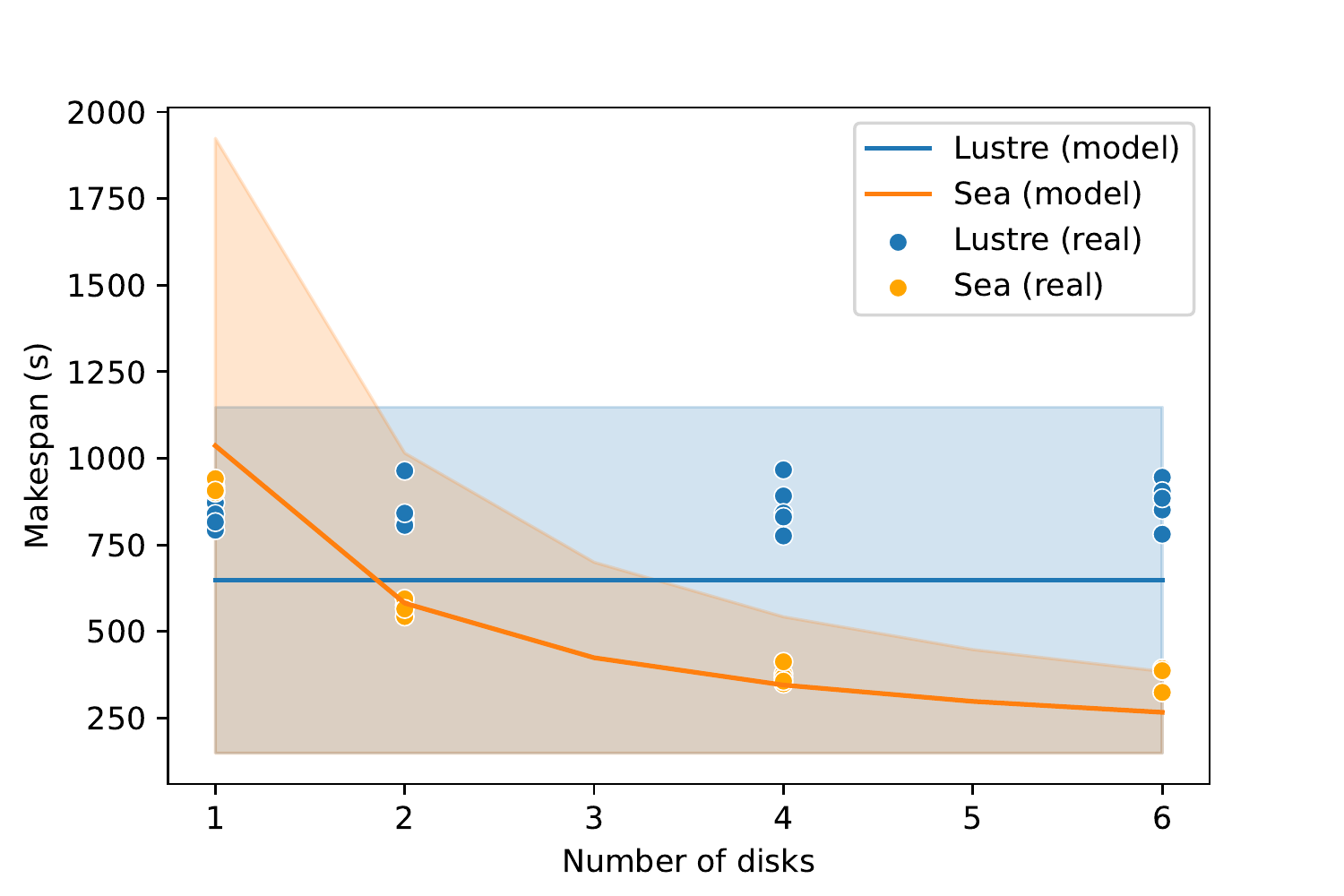}
        \caption{Experiment 2: Varying the number of disks, 5
        iterations}\label{fig:sea-comp:disks}
    \end{subfigure}
    \begin{subfigure}{\columnwidth}
        \centering
        \captionsetup{width=.85\linewidth}
        \includegraphics[width=\linewidth]{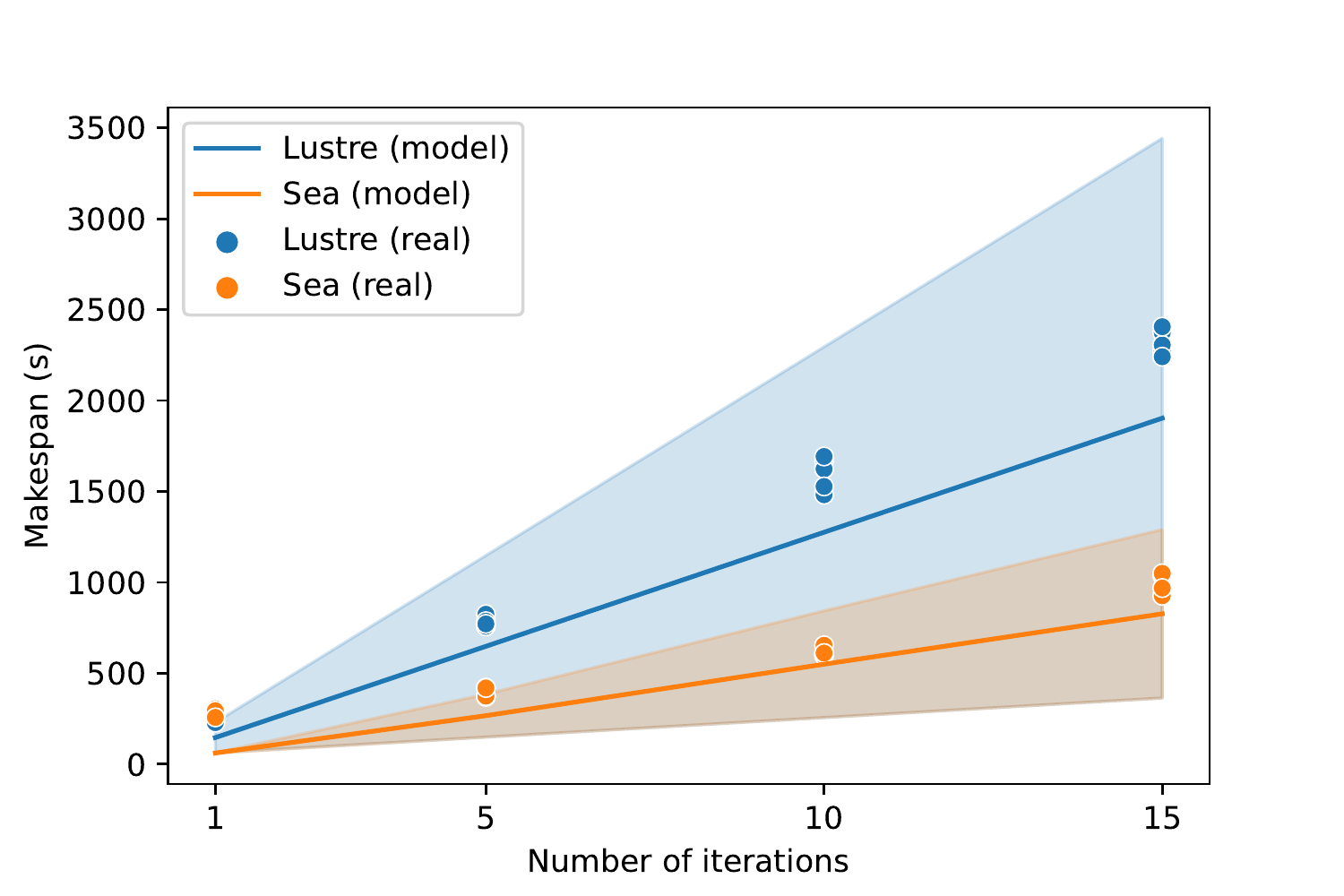}
        \caption{Experiment 3: Varying the number of
        iterations}\label{fig:sea-comp:iterations}
    \end{subfigure}
    \begin{subfigure}{\columnwidth}
        \centering
        \captionsetup{width=.85\linewidth}
        \includegraphics[width=\linewidth]{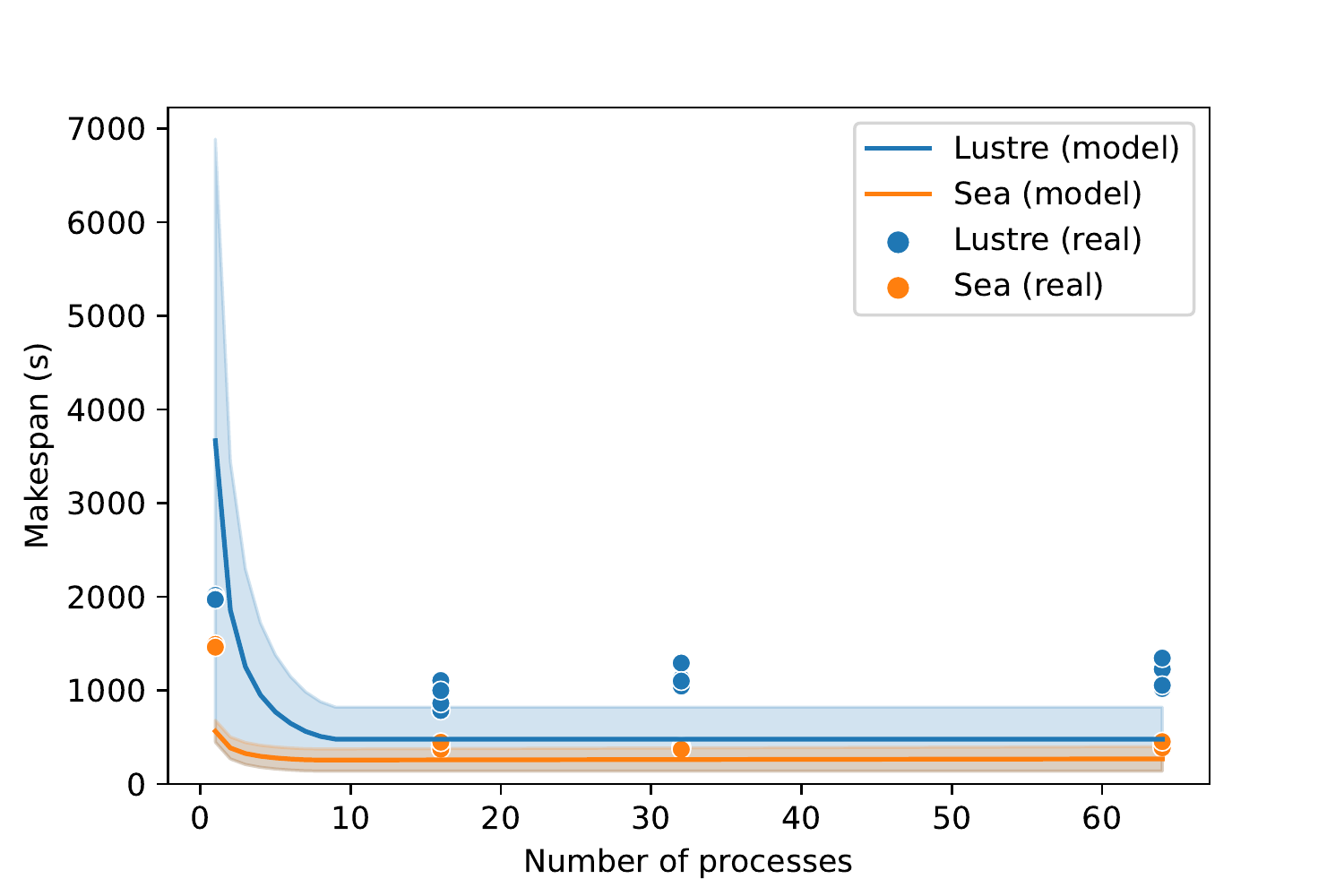}
        \caption{Experiment 4: Varying the number of parallel processes, 5
        iterations}\label{fig:sea-comp:processes}
    \end{subfigure}
    \caption{Benchmarks comparing Lustre and Sea In-memory performance under
    various conditions using the incrementation application on the Big Brain
    image. All experiments were repeated 5$\times$. Model bounds represented by
    coloured regions}
    \label{fig:sea-comp:benchmarks}
    \end{figure*}

      \subsection{Sea achieves significant speedups}

      As can be seen in Figure~\ref{fig:sea-comp:benchmarks}, Sea using
      in-memory configuration speeds up execution the majority of the
      conditions. The largest speedups observed can be seen in
      Figure~\ref{fig:sea-comp:processes} at 32 processes, where the speedup is
      nearly 3$\times$. We believe that the speedup is primarily due to the
      contention on Lustre. While each local disk has an average of 5 processes
      trying to write concurrently to it at 32 processes, each Lustre disk has around
      3 concurrent processes writing to it. Whereas one might assume this means
      Lustre would have less contention overall and should perform better, there
      are significant bandwidth differences to account for between the local
      disks and Lustre (see Table~\ref{table:sea-comp:fs}) that would result in
      improved local disk performance despite increased contention. Furthermore,
      Lustre has a centralized metadata server that is responsible for
      determining which OST is assigned to each block, guaranteeing a certain
      amount of load-balance on the storage disks. This is in contrast with the
      local disks, who do not rely on a metadata server and are selected by Sea
      via a random shuffling.

      The experimental conditions with the next largest speedup can be seen in
      Figure~\ref{fig:sea-comp:iterations} at 10 iterations with a 2.6$\times$
      speedup. It is believed that Sea performs best at 10 iterations, rather
      than 15, because that is when the majority of the writes must be made to
      local disk. While Sea is not expected to surpass the Lustre makespan,
      Lustre does have a slight performance advantage in that is able to evict
      data once it is persisted to Lustre, allowing it to make more efficient
      use of memory whenever possible. 

      When varying the number of nodes (Figure~\ref{fig:sea-comp:nodes}), we
      achieved the greatest speedup at 5 nodes, with a speedup of 2.4$\times$.
      Similar to our experiments with multiprocessing, the speedup appears
      to be due to increased contention on Lustre. However, there is one main
      difference, and that is that only Sea is experiencing increased
      contention, as the contention within the compute nodes is fixed. Both the
      model and experimental results state that the speedup is approaching a
      plateau, however, it is expected that once the number of threads writing
      to Lustre exceeds the number of OSTs (e.g., 9 nodes), we will observe an
      even greater speedup from Sea due to the increased contention on Lustre.

      As expected, our results demonstrate that with an increase in local disk,
      we can achieve greater speedups. Our results
      (Figure~\ref{fig:sea-comp:disks}) show that at 6 disks, we achieve a
      speedup of 2$\times$. This is natural, due to the fact that the most disks
      we have, the less contention there will be on any given disk. In
      particular, in our case, each disk at 6 disks should optimally only have a
      single process writing to it.

      When Sea does not provide speedups, it either performs similarly to
      Lustre, as can be seen in Figure~\ref{fig:sea-comp:nodes} at one node and
      in Figure~\ref{fig:sea-comp:iterations} at 1 iteration, or sightly slows
      down execution, as can be seen in~\ref{fig:sea-comp:disks}. As expected,
      Sea at a single iteration can at best perform similarly or slightly worse
      than Lustre. This is because all the data is read from Lustre and written
      to Lustre, operating in the same way that Lustre would with page cache.

      Sea at a single node likely performs equivalently to Lustre because
      Lustre, in this case, has very good bandwidth, having at most 6 concurrent
      writes to the whole file system. Furthermore, due to the limited number of
      concurrent writes on Lustre at any given time, Lustre needs to wait less
      for writes to flush to disk, making better use of available page cache
      space. Sea's performance, in contrast, is negatively impacted as there is
      more data being written to local storage and the combined bandwidth of the
      local storage is far less superior to that of Lustre's, due to only having
      6 disks available.

      The contention on the disk at Sea operation with only a single local disk
      is likely the cause of the decline in performance in
      Figure~\ref{fig:sea-comp:disks}. As Lustre has significantly less
      contention on the file system, it is able to exhibit a superior
      performance to Sea. As we can see, increasing the number of local disks
      improves Sea's performance beyond what Lustre can achieve.

      \subsection{Performance trends described by the model}

      Figure~\ref{fig:sea-comp:benchmarks} shows us that our model can
      describe the performance trends for all but two experiments: Experiment 3
      (Figure~\ref{fig:sea-comp:iterations}) and Experiment 4
      (Figure~\ref{fig:sea-comp:processes}). In Experiment 3, the model
      incorrectly represents the bounds for 1 iteration. Since, in this case,
      all the data should be able to fit in page cache for both Sea and Lustre,
      it is possible that memory bandwidth has been overestimated, resulting in
      incorrect model predictions.

      We believe that Lustre exceeds the model bounds in Experiment 4 due to
      increased contention on the file system. While our model predicts that the
      disk bandwidth will be the bottleneck, thus plateauing at 9 parallel
      processes per node, there are other Lustre bottlenecks that are not
      included in the model, like the metadata server. We expect that there were
      too many incoming requests to the server at 30+ parallel processes, that
      performance declined above model bounds.

    \begin{figure}

        \centering
        \includegraphics[width=\columnwidth]{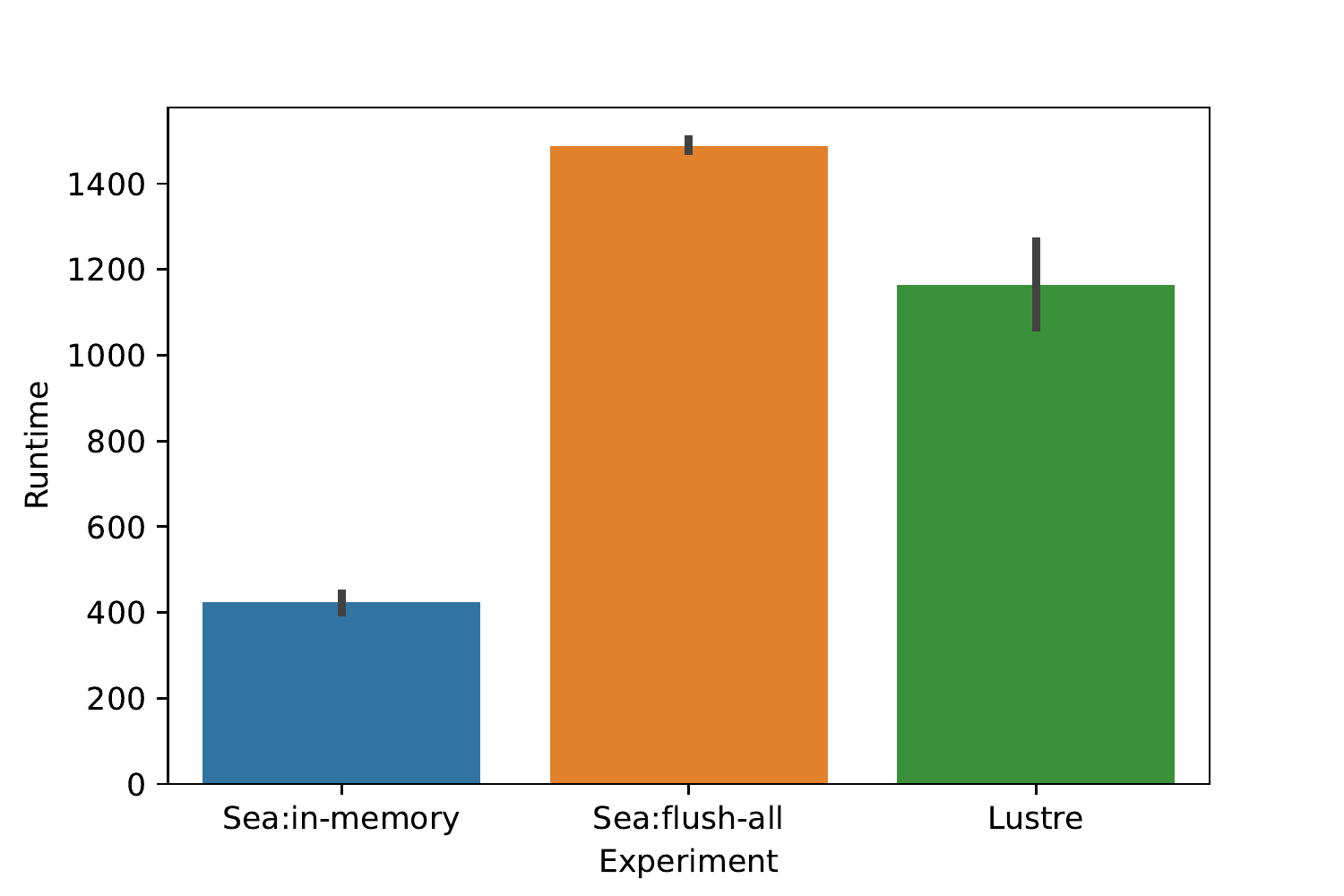}%
        \caption{Comparison of the different Sea modes with Lustre on the
        incrementation application processing 1000 Big Brain blocks with 5
        nodes, 5 iterations, 6 disks and 6 parallel processes.}
    \label{fig:sea-comp:flush}
    \end{figure}

      \subsection{Significant overheads with Flush-all mode}

      Figure~\ref{fig:sea-comp:flush} demonstrates the overhead that can be
      obtained by using flush-all when there is no compute to mask the flushing.
      Not only is Sea flush-all 3.5$\times$ slower than Sea in-memory, it is
      1.3$\times$ slower than Lustre. The reason for it being slower than Lustre
      is that Sea flush-all does not evict any files and had to copy files from
      local disk to Lustre, as a result. In contrast, Lustre does not make use
      of local disk space. It is able to evict data from memory once it is
      materialized to Lustre. Consequently, Lustre alone does not have the third
      overhead of writing the data to disk like Sea flush-all does. The
      performance loss is likely to not have been discernible if we had compute
      time that matched data transfer time, however, the performance gain might
      not have been as significant unless we were writing in parallel data that
      far exceeded page cache space as the application would not have to wait
      for data to be flushed to Lustre before proceeding.

\section{Discussion}

    \subsection{Lightweight data-management library for CLI applications}

    Through the interception of glibc calls, Sea successfully manages to
    redirect I/O to the different available storage devices on an HPC cluster.
    Since Sea is lightweight and requires minimal configuration, it preserves
    the characteristics that are important to scientific applications.

    At minimum, Sea requires the specification of a configuration file for it to
    work. A user would need to know details on the cluster storage that can be
    leveraged as well as approximate details on how much data an application
    execution can produce at any given time. Due to the simplicity of the
    configuration file, Sea maintains the ease-of-use requirement for scientific
    applications.

    Sea maintains whole files, as produced by the pipeline. In no instance does
    it modify or alter the data produced by the pipeline. Since there is no risk
    of Sea modifying the contents of the data, Sea preserves the reproducibility
    requirement.

    The flush-and-evict process may affect overall application parallelism if multiple Sea instances
    are launched on a given node. If
    only a single instance of Sea is called on a compute node, there will only
    be a single flush and evict process. However, if Sea is launched many times on a given
    node, there will be many flush and evict processes which may interfere with
    the application compute. However, our results do not indicate that
    performance is significantly impacted by the presence of a single flush and
    evict process.  

    Sea cannot be used with operating systems that are not Linux-based, and
    therefore, the use of an application with Sea has limited portability. However, Linux-based operating systems
    are ubiquitous in HPC, and therefore, Sea is not limiting for
    its intended use case. Furthermore, we provide publicly available containers~\footnote{\url{https://github.com/orgs/big-data-lab-team/packages?repo_name=Sea}}
    to simplify the usage of Sea on various other operating systems.

    \subsection{In-memory performance with Sea}
   
    Our results indicate that Sea can significantly improve the performance in
    applications executing data-intensive workflows. In all experiments, we
    observed speedups of up to 3$\times$, with the majority of cases reaching a
    2$\times$ speedup. In very few cases did the use of Sea not result in any
    speedups. In two of the three scenarios, Sea performed identically to
    Lustre, either because Sea was issuing the same amount of data transfers to
    Lustre or Lustre bandwidth far exceeded what was available locally.

    On shared cluster environments, such as those found in high-performance
    computing clusters, it is less likely that the PFS would be so underused
    that it could achieve better performance than leveraging local storage, even
    if local storage was limited. We therefore believe that it is very unlikely to
    obtain no speedups from using Sea on a production HPC cluster as long as the
    application is data intensive.
    
    For our experiments we relied on a synthetic data-intensive application,
    however, the I/O patterns exhibited by such an application do not adequately
    mimic the patterns of scientific applications. In our experiments, we
    demonstrate what the possible performance upper bound can look like and how
    it is affected by various different factors. Despite the fact that typical
    scientific workflows may not be as data-intensive, it is believed that
    scientific applications can still benefit a significant amount by limiting
    writes to a shared PFS that is experiencing traffic generated by many other
    users. Furthermore, given that scientific applications have more compute,
    Sea flush-all can be used with reduced overheads.

    \subsection{Performance boost with local storage availability}

    As expected, Sea's performance increases with the number of disks available.
    However, our results also demonstrate that it does not require that many
    disks to surpass Lustre's performance, even when Lustre is underutilized.
    Therefore, while Sea underperformed at a single disk, it is likely that this
    will be sufficient to experience speedups even in a production environment,
    where Lustre has to deal with user traffic across the cluster. This is
    important to note because it is not uncommon for HPC cluster compute nodes
    to only have a single disk available as burst buffer. However, should more
    disks be available, it would be best to include as many as possible in Sea.

    \subsection{Sea preferred when Lustre is overloaded}

    In instances where contention on Lustre exceeds that of local disks, Sea
    in-memory outperforms Lustre. In all cases, it did not take very much for
    Sea to outperform Lustre. It is expected that the resource requirements of
    scientific applications running on HPC systems could far exceed those of our
    current experiments, relying on 100s of nodes instead of just 5. Therefore,
    data-intensive scientific applications would benefit greatly from using Sea.

    When Lustre is not overloaded, there is little benefit to using Sea.
    However, we found that Sea performs similarly to Lustre in these cases.
    Since there is no real penalty to using Sea when Lustre is not overloaded,
    it is recommended to use Sea in all scenarios. Moreover, users on HPC
    clusters are unaware of what the PFS performance will be at the time when
    their experiments will be scheduled. Knowing that Sea does not incur
    significant overheads will allow users to freely execute Sea without
    hesitation. 

    \subsection{Flush when necessary and evict often}

    The results demonstrate that flushing all the data incurs significant
    overheads with a data-intensive application. These overheads can be so
    significant that Sea's performance, in these cases, can be found to be
    inferior to that of Lustre. In applications where data is shared or when
    results are required for post-processing, there is no other option than to
    flush this data. Since the majority of the overhead appears to have arisen
    from writing to and flushing from local disk, it is recommended that
    lesser-used data be evicted from Sea, freeing up space for newer data.

    There is limited benefit to Sea when flushing all the data in a
    data-intensive scenario. Sea must perform the same number of I/O operations
    as Lustre in these cases. While the application itself can proceed to
    completion faster, as it only needs to wait for all the data to be written
    to local storage, the time required for the final flush of the data can be
    quite significant, particularly when flushing from disk to Lustre.
    Therefore, we recommend that flushing all the data is reserved for more
    compute-intensive applications.

    Increase in performance from eviction is not only limited to scenarios where
    all the data is flushed. Sea also benefits from eviction when using the
    in-memory option as not all data may fit in memory and thus need to be
    written to slower local storage. Sea currently cannot handle scenarios where
    the application is attempting to access a file that is in the process of
    being moved, and as a result, we were not able to use much eviction in our
    experiments. This would be an important feature to enable in future Sea
    releases, despite the potential slowdown that may be incurred from waiting
    for the data to be materialized to Lustre.

\section{Conclusion}
    We created Sea, a lightweight open-source data-management library for
    scientific applications. With the help of glibc interception, we were able
    to create a library that maintains qualities important to scientific
    computing (i.e. ease-of-use, reproducibility, portability and parallelism).
    Our results demonstrate that Sea is quite beneficial to reducing the data
    transfers overheads, particularly when using an in-memory computing
    configuration, producing speedups of up to 3$\times$. Sea's performance is,
    however, limited when the PFS is not being heavily used.

    Experimental results demonstrate that our performance model accurately
    depicts the bounds in the majority of the cases. The model overestimated
    performance when metadata calls were heavily impacting performance. This is
    because the model neglects to account for any kind of latency. More accurate
    predictions could be obtained with a more complex model, although, due to
    Lustre's complex functioning, a simulator would likely be more appropriate
    here.

    More complex functioning of Sea, such as splitting of individual files, as
    seen with the other burst buffer file systems, may be preferential,
    particularly in maximizing cache usage. However, the more we complicate
    these libraries, the harder they are for users to use. As a result, users
    must wait until the cluster guidelines are developed detailing how a user
    should use these libraries. The boundary between ease-of-use and performance
    needs to be further explored to determine what is best for users and
    applications.


%


\section*{Acknowlegments}
Val\'erie Hayot-Sasson was funded by the Canadian Open Neuroscience Platform (CONP) Scholar Award.
This work was also supported by the Canada Foundation for Innovation as well as the Canada Research Chairs program.


\bibliographystyle{plain}
\bibliography{biblio}


\begin{IEEEbiography}[{\includegraphics[width=1in,height=1.25in,clip,keepaspectratio]{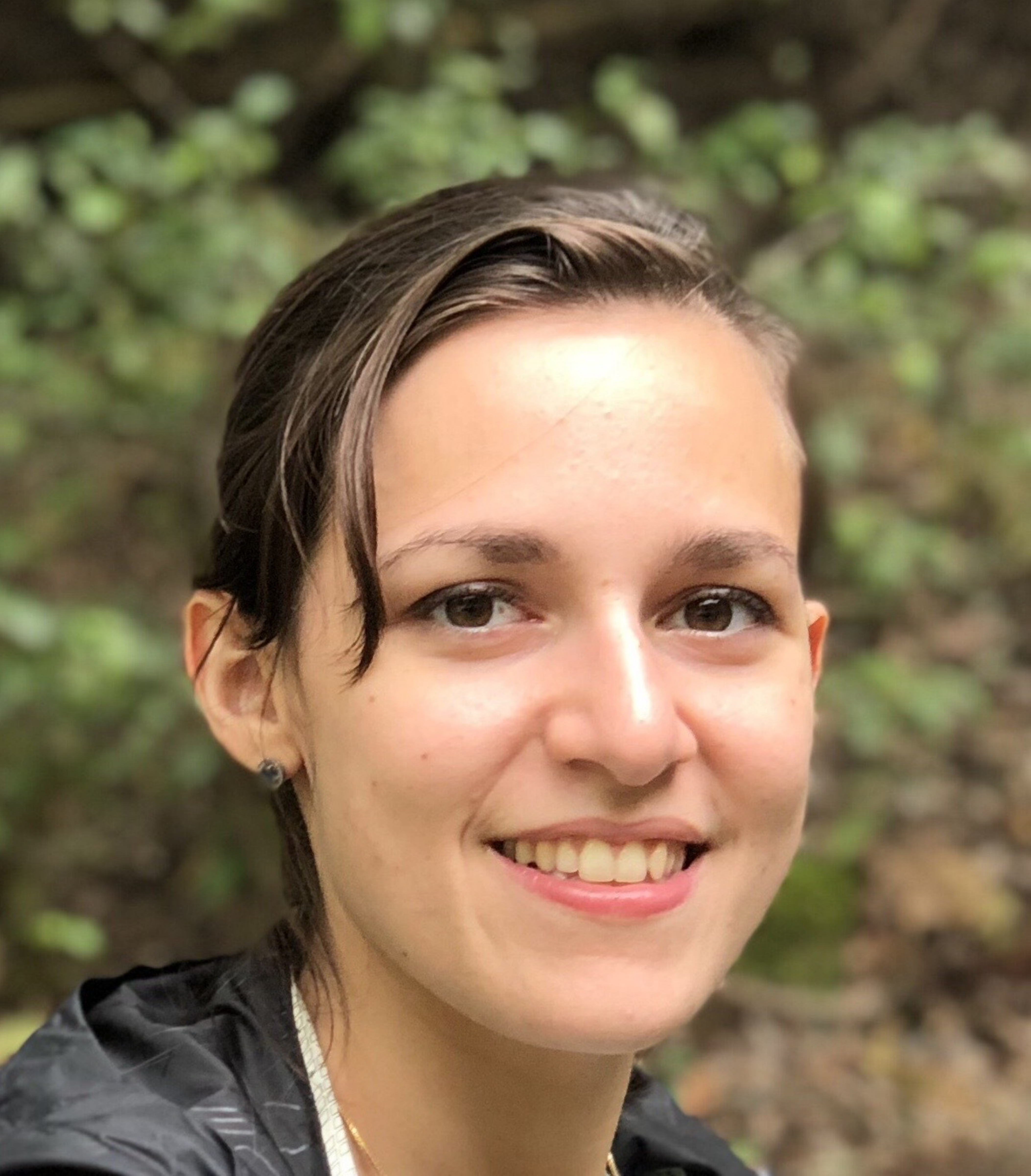}}]{Val\'erie Hayot-Sasson} is a PhD student at Concordia University in the Big Data for Neuroinformatics lab. Her research interests focus on studying the effects of data transfers on Big Data scientific application.
\end{IEEEbiography}
\begin{IEEEbiography}[{\includegraphics[width=1in,height=1.25in,clip,keepaspectratio]{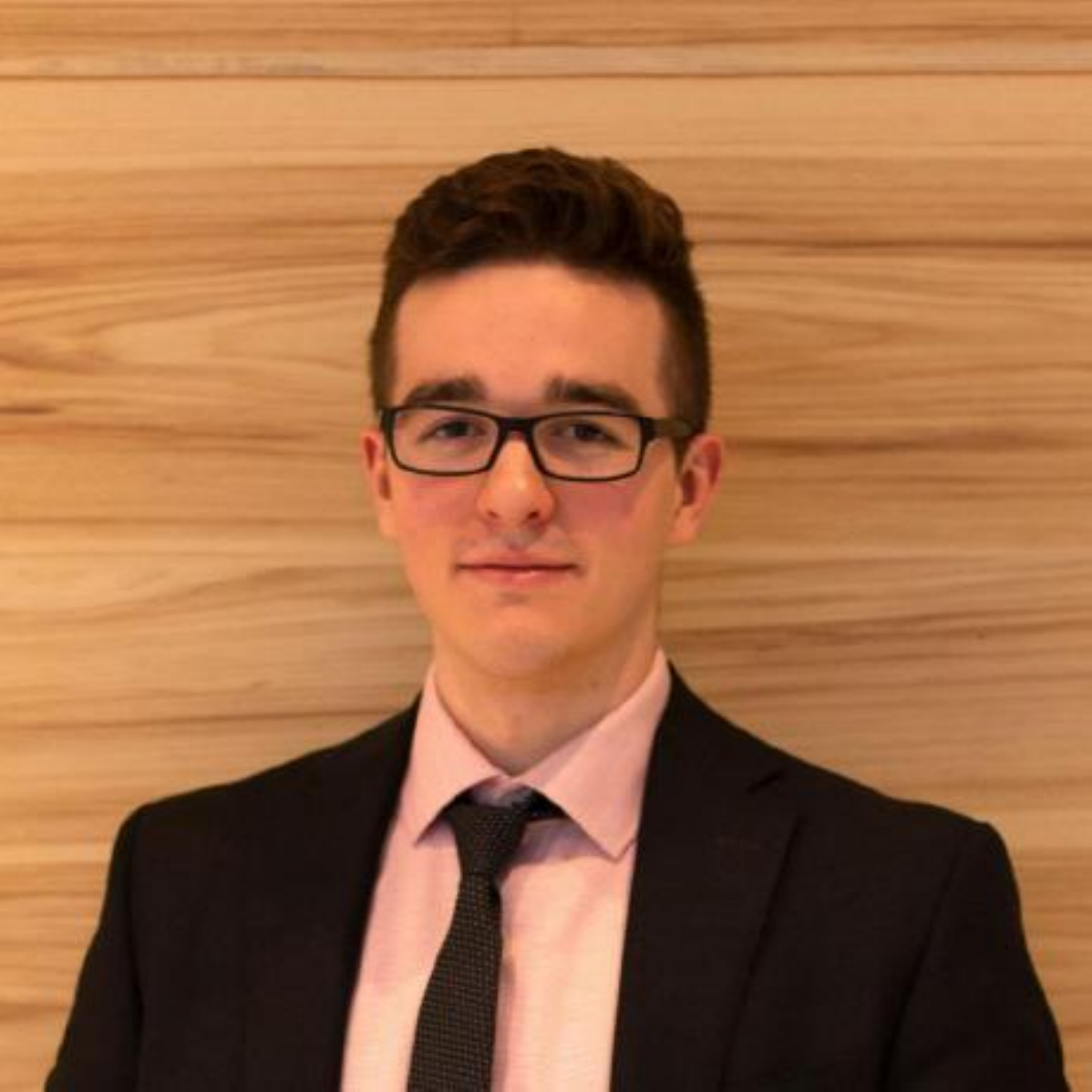}}]{Mathieu Dugr\'e} is a Ph.D. student at the Big Data Infrastructure for Neuroinformatics lab at Concordia University, Montreal, Canada. His research interests are in Big Data performance techniques for neuroimaging applications.

\end{IEEEbiography}
\begin{IEEEbiography}[{\includegraphics[width=1in,height=1.25in,clip,keepaspectratio]{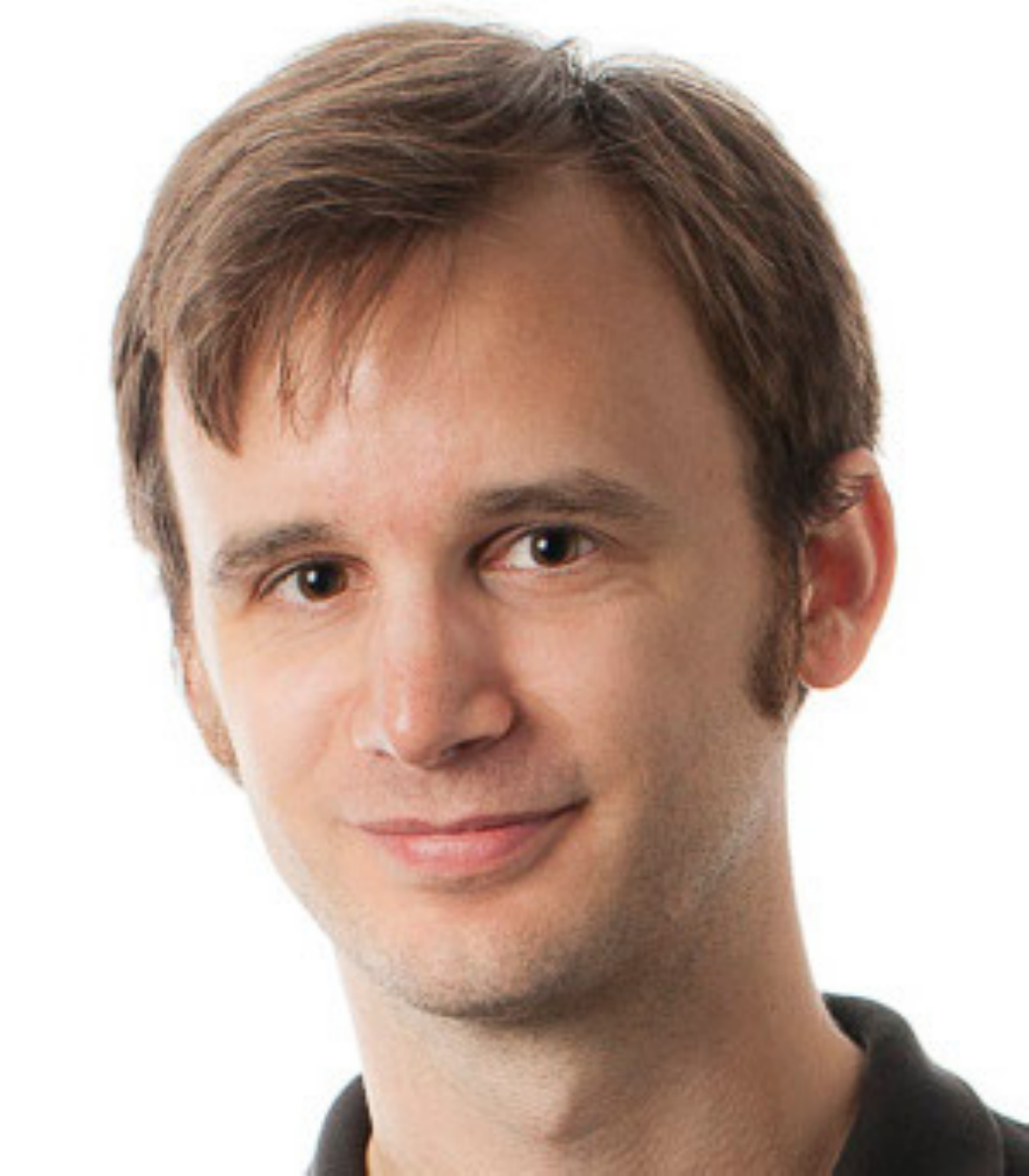}}]{Tristan Glatard} is Associate Professor in the
Department of Computer Science and Software
Engineering at Concordia University in Montreal,
and Canada Research Chair (Tier II) on Big
Data Infrastructures for Neuroinformatics. Before
that, he was research scientist at the French Na-
tional Centre for Scientific Research and Visiting
Scholar at McGill University.
\end{IEEEbiography}
\end{document}